\documentclass{PoS}

\def\la{\langle}
\def\ra{\rangle}
\def\beq{\begin{equation}}
\def\eeq{\end{equation}}
\def\be{\begin{eqnarray}}
\def\ee{\end{eqnarray}}
\def\hs{\hat{s}}
\def\htm{\hat{t}}
\def\hu{\hat{u}}

\newcommand{\f}[2]{\frac{#1}{#2}}
\newcommand{\dd}  { {\textrm d}}
\newcommand{\ba}{\begin{eqnarray*}}
\newcommand{\ea}{\end{eqnarray*}}

\title{Expected nuclear modifications and pseudorapidity asymmetry in p(d)Pb collisions at the LHC}

\ShortTitle{Pseudorapidity asymmetry in p(d)Pb collisions at the LHC}

\author{Adeola Adeluyi\\
        Center for Nuclear Research\\
        Department of Physics\\
	Kent State University\\
        E-mail: \email{aadeluy1@kent.edu}}

\author{Gergely G.~Barnaf\"oldi\\
        Center for Nuclear Research\\
        Department of Physics\\
        Kent State University\\
        E-mail: \email{bgergely@rmki.kfki.hu}}

\author{\speaker{George Fai}\\
        Center for Nuclear Research\\
        Department of Physics\\
        Kent State University\\
        E-mail: \email{gfai@kent.edu}}

\author{P\'eter L\'evai\\
        MTA KFKI RMKI Research Institute for Particle and Nuclear Physics\\
        P.O. Box 49, Budapest 1525, Hungary\\
        E-mail: \email{plevai@rmki.kfki.hu}}

\abstract{We calculate nuclear modification factors and pseudorapidity asymmetries 
in $pA$ and $dA$ collisions in a pQCD-improved parton model. With the calculations 
tuned to describe existing spectra from $pp$ collisions and asymmetric systems at 
midrapidity and large rapidities at FNAL and RHIC energies, we make predictions 
for LHC energies.}

\FullConference{High-pT Physics at LHC - Tokaj'08\\
		 March 16 - 19 2008\\
		 Tokaj, Hungary}

\begin{document}

\section{Introduction}

Asymmetric systems offer unique information about the underlying dynamics,  
not available in symmetric proton-proton or nucleus-nucleus collisions. The asymmetry
manifests itself in an asymmetric distribution of charged particles with respect to
zero rapidity (or pseudorapidity) as measured by BRAHMS~\cite{Arsene:2004cn} and 
PHOBOS~\cite{Back:2004mr}. The asymmetry can be quantified by introducing the ratio
of pseudorapidity densities at a given negative pseudorapidity relative to that at
the positive pseudorapidity of the same magnitude. This backward/forward ratio
(forward being the original direction of motion of the light partner) is referred to
as pseudorapidity asymmetry. The STAR Collaboration published pseudorapidity 
asymmetries in 200 $A$GeV $dAu$ collisions for several identified hadron species 
and total charged hadrons in the pseudorapidity intervals $|\eta| \le 0.5$ and 
$0.5 \le |\eta| \le 1.0$~\cite{Abelev:2006pp}. Asymmetries with the
backward/forward ratio above unity for transverse momenta up to $\approx 5$ GeV/c
are observed for charged pion, proton+anti-proton, and total charged hadron production 
in both rapidity regions. We anticipate that proton-lead (or deuteron-lead) data will soon 
be collected at higher energies, at the Large Hadron Collider (LHC). 

In the present study we investigate the roles of nuclear shadowing and multiple scattering 
in the generation of nuclear modifications and pseudorapidity asymmetries in a wide 
transverse-momentum range. We use the HIJING
shadowing parameterization~\cite{Li:2001xa} and the
Eskola--Paukkunen--Salgado (EPS08) nuclear parton distribution functions
(nPDFs)~\cite{Eskola:2008ca}. While the former has been applied widely, 
the latter was not available at the time of similar
earlier studies. We present pseudorapidity asymmetries 
for $pBe$ at $30.7$ GeV (Fermilab) and $dAu$ at $200$ $A$GeV (RHIC), where data are
available from the E706 experiment~\cite{Apanasevich:2002wt},
PHENIX~\cite{PHENIXdAu}, and STAR~\cite{Abelev:2006pp}. We then make predictions 
for $dPb$ at $8.8$ $A$TeV (LHC). 

\section{Model framework}

The invariant cross section for the production of 
final hadron $h$ from the collision of nucleus $A$ and nucleus $B$ 
($A+B \!\to\! h+X$), can be written as
\begin{eqnarray}\label{eq:pdA}
E_h\f{\dd ^3 \sigma_{AB}^{h}}{\dd^3 p} =
\sum_{\!\!abcd}\!\int\!\! \dd^2b\ \dd^2r\ t_A(b) t_B(|\vec{b}-\vec{r}|)\dd{x_a} \dd{x_b}
 \dd{\vec k}_{Ta} \, \dd{\vec k}_{Tb} \, \dd{z_c} \,
f_{\!a/A}(x_a,\!{\vec k}_{Ta},Q^2)\ \nonumber \\
f_{\!b/B}(x_b,\!{\vec k}_{Tb},Q^2)\
\f{\dd\sigma(ab\!\to\!cd)}{\dd\htm}\,
\frac{D_{h/c}({z_c},\!Q_f^2)}{\pi z_c^2}
\hs \, \delta(\hs\!+\!\htm\!+\!\hu)   \,\,  ,
\end{eqnarray}
where all quantities have their usual meaning~\cite{Zhang:2001ce}.

The collinear parton distribution functions (PDFs) are generalized to include a transverse
momentum degree of freedom, ${\vec k}_T$, as required by the uncertainty principle. This
can be formally implemented in terms of unintegrated PDFs~\cite{Collins:2007ph,Czech:2005vy}. 
To avoid some of the complications associated with using unintegrated
PDFs, we resort to a simple factorized approximation, where  
the $k_T$-broadened parton distribution in the nucleon is written as
\begin{equation}
f_{a/N}(x,{\vec k}_{T},Q^2) \longrightarrow  \ g({\vec k}_T) \cdot
f_{a/N}(x,Q^2),  \,\,\,\,\, 
g({\vec k}_T)  =  \frac{\exp(-k_T^2/\langle k_T^2 \rangle_{pp})} {\pi \langle k_T^2 \rangle_{pp}} \,\, ,
\end{equation} 
with  $f_{a/N}(x,Q^2)$ denoting the standard collinear PDF in the
nucleon, and 
$\langle k_T^2 \rangle_{pp}$ is the two-dimensional width of the transverse-momentum
distribution in the proton,
related to the magnitude of the average transverse momentum of a 
parton by $\langle k_T^2 \rangle_{pp} = 4 \langle k_T \rangle_{pp}^2/\pi$. 

Most shadowing parameterizations include at least some of the effects
of multiple scattering in the nuclear medium, while the HIJING parameterization
(as we discuss further in Sec.~\ref{shad_multiscatt}) needs to be augmented
with modeling nuclear multiscattering. For this purpose, in $pA$ collisions
we use a broadening of the width of the transverse momentum distribution 
according to 
\begin{equation}\label{eq:broad}
\langle k_T^2\rangle_{pA} = \langle k_T^2\rangle_{pp} + C\ h_{pA}(b) \,\, ,
\end{equation}
where $\langle k_T^2\rangle_{pp} $ is the width already present in proton-proton 
collisions, $h_{pA}(b)$ is the number of effective nucleon-nucleon ($NN$) collisions as a 
function of nucleon impact parameter $b$, and $C$ is the average increase in width 
per $NN$ collision. 

The collinear nPDFs $f_{a/A}(x,Q^2)$ are 
expressible as convolutions of nucleonic parton distribution functions (PDFs)
$f_{a/N}(x,Q^2)$ and a shadowing function $S_{a/A}(x,Q^2)$ which encodes the 
nuclear modifications of parton distributions.  We use the MRST2001
next-to-leading order (NLO) PDFs~\cite{Martin:2001es} for the nucleon 
parton distributions and for the shadowing function we employ both 
the EPS08 shadowing routine~\cite{Eskola:2008ca} and HIJING~\cite{Li:2001xa}.
(Other nPDFs, like FGS~\cite{Frankfurt:2003zd}, HKN~\cite{Shad_HKN}, 
and the earlier EKS~\cite{Eskola:1998df}, are used elsewhere
to calculate pseudorapidity asymmetries~\cite{Adeluyi:2008qk}.) 
For the final hadron fragmentation we utilize the fragmentation functions
in the AKK set~\cite{Albino:2005me}. The factorization and 
fragmentation scales are from the best-fit results obtained in
Ref.~\cite{Levai:2006yd}. 
We obtain the density distribution of the deuteron from the Hulthen 
wave function~\cite{Hulthen1957} (as in Ref.~\cite{Kharzeev:2002ei}), while a 
Woods-Saxon density distribution is used for gold and lead with parameters from
Ref.~\cite{DeJager:1974dg}.

\section{Nuclear modifications and pseudorapidity asymmetry}

The nuclear modification factor, $R^h_{AB}(p_T, \eta)$, and
pseudorapidity asymmetry, $Y^h_{Asym}(p_T)$, can be 
defined for any produced hadron species $h$ as
\begin{equation}
R^h_{AB}(p_T, \eta) = \frac{1}{\langle N_{bin}\rangle} \cdot
\frac{E_h \dd^3\sigma_{AB}^{h}/\dd^3 p |_{\eta}}
{E_h \dd^3\sigma_{pp}^{h}/\dd^3 p |_{\eta}},   \,\,\,\,\, 
Y^h_{Asym}(p_T) = \left. E_h\f{\dd ^3 \sigma_{AB}^{h}}{\dd^3 p} \right|_{-\eta} 
 \left/ 
\left. E_h\f{\dd ^3 \sigma_{AB}^{h}}{\dd^3 p} \right|_{\eta} \right. \,\, ,
\label{rdau}
\end{equation}
where $\la N_{bin} \ra$ is the average number of binary collisions.

Let us consider the (double) ratio of the  
forward and backward nuclear modification factors in $dAu$
collisions for species $h$:
\begin{eqnarray}
R^h_{\eta}(p_T) = \frac{R^h_{dAu}(p_T,-\eta)}{R^h_{dAu}(p_T,\eta)}=   
\frac{E_h \dd^3\sigma_{dAu}^{h}/\dd^3 p |_{-\eta}}
{E_h \dd^3\sigma_{pp}^{h}/\dd^3 p |_{-\eta}} \left/  
\frac{E_h \dd^3\sigma_{dAu}^{h}/\dd^3 p |_{\eta}}
{E_h \dd^3\sigma_{pp}^{h}/\dd^3 p |_{\eta}} \right. \,\, . 
\label{y-r:eq}
\end{eqnarray}
As discussed in Ref.~\cite{Barnafoldi:2008rb}, the $pp$ 
rapidity distribution is symmetric around $y=0$ and thus cancels.
We therefore obtain: 
\begin{equation}
Y^h_{Asym}(p_T)= R^h_{\eta}(p_T)=\frac{R^h_{dAu}(p_T,-\eta)}{R^h_{dAu}(p_T,\eta)} \,\, .
\end{equation}

\section{Nuclear shadowing and multiple scattering}
\label{shad_multiscatt}

We expect that particle production in $pA$ and $dA$ collisions will be 
different in the forward and backward directions. This is because the
respective partons have different momentum fractions (shadowing differences)
and because the forward-going parton has to traverse a large amount of matter.
In this Section we therefore 
study the interplay of shadowing and multiple scattering in more detail 
using the HIJING parameterization. 

\begin{figure}[!htb]
\begin{center} 
\includegraphics[width=11.5cm, height=6.4cm, angle=0]{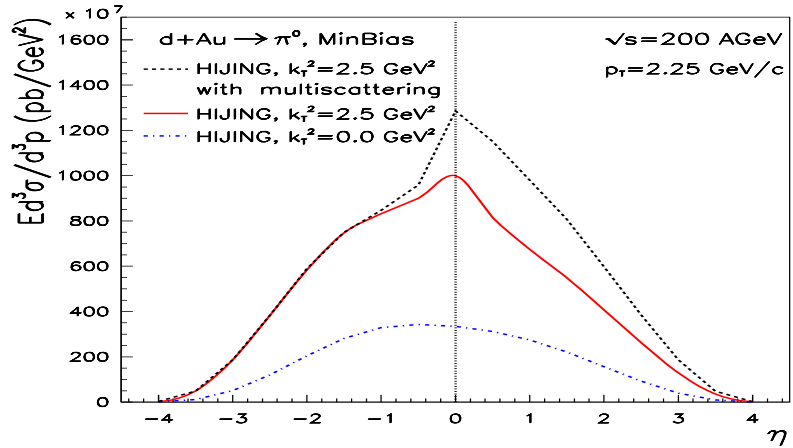}
\end{center}
\caption[...]{(Color Online) Pseudorapidity distribution for 
minimum bias neutral pion production from $dAu$ collisions at 200 $A$GeV.
The solid line represents HIJING shadowing with intrinsic $k_T$ in the
proton, the dashed line is HIJING with intrinsic $k_T$ and multiple 
scattering, while the dot-dashed curve is obtained by turning off
any transverse momentum.}
\label{fig:multiscatt}
\end{figure}

In Fig.~\ref{fig:multiscatt} we display the pseudorapidity distribution for 
neutral pion production from $dAu$ collisions at a fixed $p_T$ ($= 2.25$ GeV/c). 
The parameters of eq.~(\ref{eq:broad}) are unchanged from our previous 
studies at midrapidity~\cite{Levai:2003at,Levai:2006yd}, and 
we have chosen a transverse momentum value comparable to $\la k_T \ra_{pp}$,
where the various effects are clearly displayed. 

The results of our study can be summarised as follows: In the absence
of shadowing, the distribution is symmetric around midrapidity. With
shadowing only, an asymmetry develops: the yield at a fixed negative
pseudorapidity, say $\eta = -2$ ($Au$ side or ``backward'') 
is higher than at the corresponding positive pseudorapidity ($y=2$, $d$ side,  
or ``forward''). Thus the pseudorapidity asymmetry, $Y_{Asym}$, is greater 
than unity in this case. The inclusion of the intrinsic transverse momentum 
in the proton, while increasing the yields on both sides, does not
affect the forward/backward symmetry.
When multiple scattering is included, the yield shows a stronger increase in 
the forward direction, leading to $Y_{Asym}$ less than unity.
Thus we can say that shadowing suppresses the yield more on the $d$
side (forward) relative to a symmetric collision,
while the multiple scattering contribution is understandably large in the 
forward direction.
 
We have carried out similar studies at higher $p_T$ values. When $k_T
<< p_T$, then $k_T$
effects become naturally smaller. Shadowing effects also become smaller as
the antishadowing region of the HIJING parameterization is approached. Thus,
at $p_T \gtrsim 15$ GeV/c the influence of multiple scattering and intrinsic
$k_T$ become negligible. These phenomena are most important at intermediate $p_T$ 
values, 2 GeV/c $\lesssim p_T \lesssim$ 8 GeV/c at RHIC. It is interesting to note
that a similar transverse-momentum region is sensitive to nuclear effects 
at lower (e.g. CERN SPS) energies, due to the $\sim \log(\sqrt{s})$ scaling of the 
Cronin peak~\cite{e706,Zielinski,Barnafoldi:2006qm}. 

\section{Pseudorapidity asymmetry}
\label{etasmy}

\subsection{Asymmetry in $pBe$ collisions at $30.7$ GeV}
\label{asympBe}

\begin{figure}[!h]
\begin{center} 
\includegraphics[width=7.5cm, height=9.5cm, angle=270]{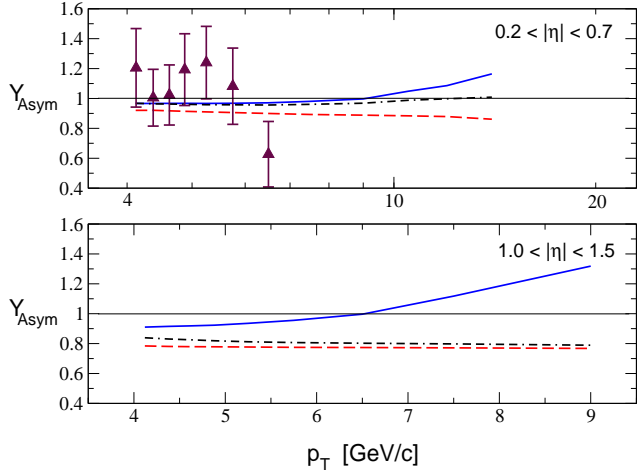}
\end{center}
\caption[...]{(Color Online) Pseudorapidity asymmetry,
$Y_{Asym}$ for $p+Be \rightarrow \pi{^0}+X$ at $0.2 < |\eta| < 0.7$ (top)
and  $1.0 < |\eta| < 1.5$ (bottom). The solid line represents the
EPS08 nPDFs, while the dashed line is obtained from HIJING with 
the inclusion of multiscattering. The dot-dashed line 
corresponds to HIJING without multiscattering, and filled triangles denote 
the E706 data~\cite{Apanasevich:2002wt}.}
\label{fig:asyfnal}
\end{figure}

The pseudorapidity asymmetry for the rapidity interval
$0.2 < |\eta| < 0.7$ is shown in the upper panel of Fig.~\ref{fig:asyfnal} 
compared with the E706 data~\cite{Apanasevich:2002wt}.
Both EPS08 with $k_T$ and HIJING 
without multiple scattering give very small asymmetries and the data are
also consistent with $Y_{Asym}=1$ for low $p_T$. The HIJING parameterization
with multiple scattering yields somewhat lower values at all transverse momenta.
This is due to multiple scattering moderately increasing 
the yield on the $p$-side relative to that of the $Be$-side. In view of the rather 
large error bars, all three sets are in reasonable agreement with the data. 
The lower panel is our prediction for the interval $1.0 < |\eta| < 1.5$. 
The calculated effects are larger, but with similar structure to
those at lower $\eta$. 


 \subsection{Asymmetry in $dAu$ collisions at $200$ $A$GeV}
\label{asymdAu}

\begin{figure}[!h]
\begin{center} 
\includegraphics[width=8.1cm, height=13.5cm, angle=270]{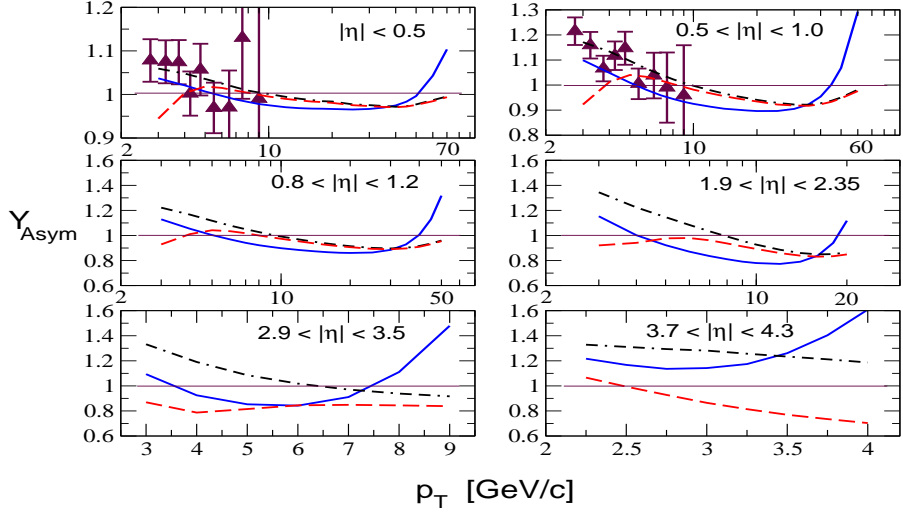}
\end{center}
\caption[...]{(Color Online) Pseudorapidity asymmetry,
$Y_{Asym}$ for $d+Au \rightarrow \pi{^0}+X$ at different 
pseudorapidity intervals. The solid line represents the
EPS08 nPDFs, while the dashed line is obtained using HIJING shadowing with 
the inclusion of multiscattering. The dot-dashed line 
corresponds to HIJING without multiscattering, and filled triangles denote 
the STAR data~\cite{Abelev:2006pp}.}
\label{fig:asyrhic}
\end{figure}

Figure~\ref{fig:asyrhic} shows the pseudorapidity asymmetry for $\pi^0$ 
production from $dAu$ collisions at RHIC, for different pseudorapidity intervals. 
The two uppermost panels are our results for the asymmetry at $|\eta| < 0.5$ 
and $0.5 < |\eta| < 1.0$ compared with the STAR data~\cite{Abelev:2006pp}. For
$p_T > 4.0$ GeV/c, the agreement with data is quite good for all three sets.
At lower $p_T$, multiple scattering increases the calculated yield mostly in the 
forward direction as discussed in Sec.~\ref{shad_multiscatt}, leading to calculated
asymmetries below unity. At very high $p_T$ we observe a divergence in 
the model predictions.
The lower four panels are our predictions for the asymmetry as pseudorapidity 
increases. The first three of these correspond to the BRAHMS
pseudorapidity intervals~\cite{Arsene:2004ux}. The general trend is that the
asymmetry becomes larger as $\eta$ increases. This mainly arises
from the strong shadowing in the larger nucleus at lower $x$ values. Also, 
since increasing $\eta$ leads to decreasing accessible $p_T$ 
due to phase space constraints, the effects of multiple scattering become 
more pronounced. 
\subsection{Asymmetry in $dPb$ collisions at $8.8$ $A$TeV}
\label{asymdPb}

\begin{figure}[!h]
\begin{center} 
\includegraphics[width=7.5cm, height=8.5cm, angle=270]{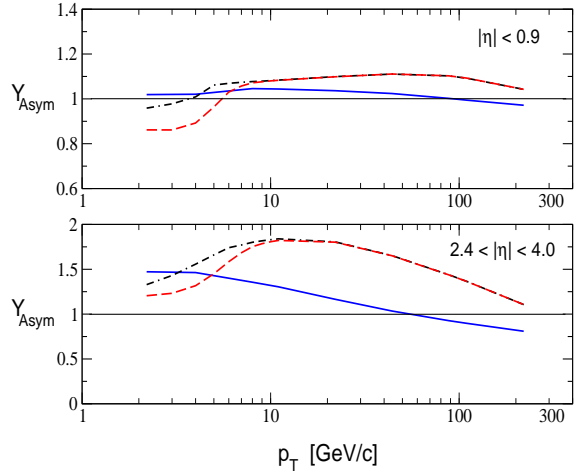}
\end{center}
\caption[...]{(Color Online) Predicted pseudorapidity asymmetry,
$Y_{Asym}$ for $d+Pb \rightarrow \pi{^0}+X$ at $\sqrt{s} = 8.8$ $A$TeV
for $|\eta| < 0.9$ and $2.4 < |\eta| < 4.0$. The solid line represents the
EPS08 nPDFs, while the dashed line is obtained from HIJING with 
the inclusion of multiscattering. The dot-dashed line 
corresponds to HIJING without multiscattering.}
\label{fig:asylhc}
\end{figure}

Let us now turn to our predictions for the pseudorapidity asymmetry at the
LHC energy of $8.8$ $A$TeV. The calculated results are displayed in 
Fig.~\ref{fig:asylhc}, where the upper panel is for the interval $|\eta| < 0.9$ 
and the lower panel is for $2.4 < |\eta| < 4.0$. These 
intervals correspond to acceptance in the central detector and in the muon 
arm, respectively, of the ALICE experiment~\cite{Alessandro:2006yt}. All three 
sets predict minimal asymmetry of the order of a few percent for the interval 
$|\eta| < 0.9$. 
As we move to higher $\eta$, the predicted asymmetry becomes more significant. 
As can be seen in the lower panel of Fig.~\ref{fig:asylhc}, both EPS08 and 
HIJING predict substantial asymmetry. According to the model, an LHC measurement 
of pseudorapidity asymmetry at around $p_T \approx 40$~GeV/c could discriminate 
between the EPS08 and HIJING shadowing prescriptions.

At the present level, neither model variant gives agreement with all aspects of 
the data: in an earlier calculation we have found that shadowing 
parameterizations which do not need to be augmented by a multiple scattering 
prescription~\cite{Eskola:1998df,Shad_HKN,Frankfurt:2003zd} have difficulty 
describing central-to-peripheral ratios at forward rapidity~\cite{Adeluyi:2008qk}. 
We have checked that this also holds for the EPS08 nPDFs. On the other hand,
the HIJING parameterization with multiscattering yields pseudorapidity 
asymmetries below unity at low transverse momenta. 
\section{Conclusion}
\label{concl}

The mechanisms for particle production in asymmetric collisions leads 
to observable asymmetries in the pseudorapidity distribution of the 
produced particles at some collision energies and transverse momenta.
We have considered the effects of nuclear shadowing 
and multiple scattering on pseudorapidity asymmetry for three asymmetric 
systems: $pBe$, $dAu$, and $dPb$ in a wide c.m. energy range from
FERMILAB up to LHC energies.

Overall, the calculated asymmetries are in reasonable agreement with available 
experimental data. Intrinsic transverse momentum in the nucleon is
seen to be important at low $p_T$. Multiple scattering 
increases the yield in the ``forward'' or positive pseudorapidity region,
thus leading to a tendency for asymmetries less than unity at low $p_T$
in a scheme explicitly relying on multiple scattering,
at variance with the data. An LHC measurement 
of pseudorapidity asymmetry in $dPb$ collisions may be able to distinguish 
between the EPS08 and HIJING nPDFs at high pseudorapidity and high $p_T$.

A major constraint in assessing pseudorapidity asymmetries is the limited 
availability of data for direct comparison with theoretical calculations. 
More data in asymmetric light-on-heavy collisions separated with respect
to positive and negative pseudorapidities are needed
to judge calculated pseudorapidity asymmetries. At RHIC, the data from
the high-statistic $dAu$ Run 8 can be used to provide such a large data sample.

\section{Acknowledgments}
\label{ack}

This work was supported in part by Hungarian OTKA PD73596,
T047050, NK62044, and IN71374, by the U.S. Department of Energy under
grant U.S. DOE DE-FG02-86ER40251, and jointly by the U.S. and Hungary under
MTA-NSF-OTKA OISE-0435701.

%

\end{document}